\documentclass[12pt,a4paper,final]{article}
\author{María Cecilia Tomasini}
\usepackage[utf8]{inputenc}
\usepackage{amsmath}
\usepackage{amsfonts}
\usepackage{amssymb}
\usepackage{graphicx}
\usepackage{float}
\title{Síntesis}

\usepackage{subfig}
\usepackage{graphicx}

\usepackage{color}

\def\ai{\'{\i}}

\def\lp{\left(}
\def\rp{\right)}
\def\lbr{\left[}
\def\rbr{\right]}
\def\rk{\right\}}
\def\lk{\left\{}

\def\be{\begin{equation}}
\def\ee{\end{equation}}
\def\ba{\begin{array}}
\def\ea{\end{array}}
\def\bea{\begin{eqnarray}}
\def\eea{\end{eqnarray}}

\def\p{\partial}

\def\t{\theta}
\def\T{\Theta}
\def\lb{\label}
\def\d{\delta}

\def\r{\rho}

\def\n{\nu}
\def\a{\alpha}
\def\b{\beta}
\def\g{\gamma}
\def\G{\Gamma}
\def\s{\sigma}

\def\om{\omega}
\def\l{\lambda}
\def\P{\Phi}

\def\s{\sigma}

\def\e{\epsilon}

\def\k{\kappa}
\def\z{\zeta}

\def\be{\begin{equation}}
\def\ee{\end{equation}}
\def\ba{\begin{array}}
\def\ea{\end{array}}
\def\bea{\begin{eqnarray}}
\def\eea{\end{eqnarray}}

\begin{document}

\title{\vspace{-1.5cm} \bf Self-force on an arbitrarily coupled scalar charge in cylindrical  thin-shell spacetimes}
\author{C. Tomasini\footnote{e-mail: ctomasini@df.uba.ar}, E. Rub\ai n de Celis\footnote{e-mail: erdec@df.uba.ar}, C. Simeone\footnote{e-mail: csimeone@df.uba.ar}\\
{\footnotesize Departamento de F\ai sica, Facultad de Ciencias Exactas y Naturales, Universidad de}\\
{\footnotesize Buenos Aires and IFIBA, CONICET, Ciudad Universitaria, Buenos Aires 1428, Argentina.}}
\date{\small \today}

\maketitle
\vspace{-0.6cm} 
\begin{abstract}

{\footnotesize We consider the arbitrarily coupled field and self-force of a static massless scalar charge in cylindrical spacetimes with one or two asymptotic regions, with the only matter content concentrated in a thin-shell characterized by the trace of the extrinsic curvature jump $\k$.
The self-force is studied numerically and analytically in terms of the curvature coupling $\xi$.
We found the critical values $\xi_c^{(n)} = n/\lp \r(r_s)\,\k \rp$, with $n \in \mathbb{N}$ and $\r(r_s)$ the metric's profile function at the position of the shell, for which the scalar field is divergent in the background configuration. 
The pathological behavior 
is removed by restricting the coupling to a domain of stability.
The coupling has a significant influence over the self-force at the vicinities of the shell, and we identified $\xi=1/4$ as the value for which the scalar force changes sign at a neighborhood of $r_s$; if $\k(1-4\xi)>0$ the shell acts repulsively as an effective potential barrier, while if $\k(1-4\xi)<0$ it attracts the charge as a potential well.
The sign of the asymptotic self-force only depends on whether there is an angle deficit or not on the external region where the charge is placed; conical asymptotics produce a leading attractive force,
while Minkowski regions produce a repulsive asymptotic self-force.}

\vspace{0.6cm} 

\noindent 
PACS number(s): 4.20.-q, 04.20.Gz, 04.40.-b, 41.20.Cv\\
Keywords: General Relativity; scalar self force; thin shells; cylindrical spacetimes; wormholes

\end{abstract}

\section{Introduction}
\lb{s1}

A field theory including a complex scalar field coupled to a gauge field predicts that spontaneous symmetry breaking can lead to cylindrical topological defects known as local or gauge cosmic strings \cite{vilenkin}. The gravitational effects of such objects have been considered of  particular physical interest within the study of structure formation in the early Universe, since they could have acted as possible ``seeds'' for density fluctuations \cite{vilenkin2,turok,sato}; besides, strings could be, in principle, observed by gravitational lensing effects.  Though in the present day theoretical framework cosmic strings are not considered as the main source of the primordial cosmological matter fluctuations,
they are still taken as a possible secondary source of fluctuations \cite{danos}; this motivates a recently renewed interest in their study.  

Local or gauge strings are characterized by having an energy-momentum tensor whose only non null components are $T_0^{\, 0}=T_z^{\, z}$. The metric around a gauge string was first calculated by Vilenkin \cite{vilenkin3} in the linear approximation of general relativity, using a Dirac delta to approximate the radial distribution of the energy-momentum tensor for a cosmic string along the $z$ axis. Thus, it was proposed $
{\tilde T}_\mu^{\, \nu} 
= \delta (x)\delta (y) \,\mbox{diag}(\mu,0,0,\mu),$
 where $\mu$ is the linear mass density, which is determined by the energy scale at which the symmetry breaking process took place. Under this assumption, and working up to first order in $G\mu$, Vilenkin obtained a spacetime around the string which is flat but presents  a deficit angle $\Delta \varphi =8\pi G\mu$. 
Some years later, Hiscock \cite{hiscock}, motivated by the possibility of theories which may lead to  large values of $G\mu$ (i. e. $G\mu\sim 1$),                                                                                                                                                                                                                                     considered a thick  cylinder of radius $a$ with uniform tension and linear mass density, whose energy-momentum tensor is $
 T_\mu^{\, \nu}(x,y)=\mbox{diag}(\mu,0,0,\mu)\theta(r-a)/ a^2,$
and solved the full Einstein equations in the interior and matched the resulting static metric with the vacuum solution for the exterior; no matter layer was assumed to exist at the radius $a$, so that the adopted matching conditions imposed the continuity of both the metric and the extrinsic curvature.  His work showed that Vilenkin's results for the metric outside the string core were actually valid to  all orders in $G\mu$, that is, the conical geometry was not an artifact introduced by the linearized approximation, but an essential feature of gauge strings. Since the corresponding metric has $g_{00}=1$, i.e. the Newtonian potential is null, non charged rest particles would not be affected by the gauge string gravity. However, strings moving at relativistic speeds would induce waves which could eventually lead to observable matter density variations.  

Things change considerably, however, when charged particles are taken into account. The local flatness of a conical background manifold does not imply zero force on a rest charge. A vanishing force requires a field symmetric around the charge, and this is not possible when a deficit angle exists. While the field equations of a given point source are locally those of Minkowski spacetime, the globally correct solution in a conical geometry is not symmetric around the source, and as a result a self-force appears on the charged particle \cite{linet}. Moreover, as shown in \cite{eoc1,eoc2,emilio} for electrically charged point particles, the behavior of such force in relation with the position of the charge allows to distinguish between two locally identical geometries which differ in their global topological aspects. This provides a sort of tool for detecting wormholes of the thin-shell kind, whose geometries are locally the same of those of the type which they would connect \cite{visser}. These wormholes are characterized by connecting two exterior geometries of a given class by a throat, which is a minimal area or circumference surface (see below), where a matter layer of negligible thickness is placed. Hence, for example, by studying the force on a static charged particle an observer could determine if the background is that of  a gauge cosmic string or that of a thin-shell wormhole connecting two conical spacetimes. Also, in the case of topologically trivial spacetimes, an analogous analysis would allow to distinguish different interior geometries by studying the self-force on a charge placed at the exterior region beyond a matter shell \cite{emilio,drivas,isomaya}. 
The problem of the self-force on a scalar point charge appears as a natural but not at all trivial extension of previous work. 

Scalar fields are widely used in cosmological models to explain the early Universe and the present accelerated expansion; non-minimally coupled to gravity scalar fields are included in inflationary models \cite{marsh}, while the unknown nature of dark matter is sometimes described by classical light or massless scalar fields \cite{diana,diana2}. 
Besides the physical interest of scalar charges and the self-force problem \cite{wisemann,burko1,warburton,harte,bini,thornburg}, a central aspect motivating a detailed analysis is the different form in which the scalar field couples to gravity: while the field of an electric charge is continuous across a non charged matter shell separating two regions of a spacetime, this is not the case for a scalar charge, and matter shells induce a sort of new sources for the corresponding scalar field; this necessarily translates to an, in principle, different behaviour of the self-force. 
Some peculiarities due to non minimal coupling of the scalar and the gravitational fields were pointed out in \cite{wisemann,zelnikov,burko,pfenning,bezerra1,popov}.
In particular, some authors who studied the scalar self-force and propagation of a scalar field in a fixed background found that the stability regions of solutions for the non minimally coupled wave equation are restricted by the value of the coupling constant.
An anomalous divergence was first realized by Bezerra and Khusnutdinov \cite{bezerra} who reported an infinite scalar self-force for some critical values of the coupling constant in a class of spherically symmetric wormhole spacetimes. For example, in a spherical wormhole with an infinitely short throat of the thin-shell kind they found critical values at $\xi=n/4$, for $n \in \mathbb{N}$. Continuing to previous work, Taylor \cite{taylor1} notice that solutions to the scalar wave equation sourced by a point charge are unstable in the Ellis wormhole for the discrete set of couplings $\xi=n^2/2$.
The pathological behavior was removed by restricting the coupling constant to the domain of stability where no poles are encountered \cite{taylor2}.
In the present work we find that a similar feature occurs in cylindrical wormholes and also in trivial spacetimes.

In this article we study the self-force acting on a static point like scalar charge in a background constructed by cutting and pasting two cylindrical manifolds with different deficit angle, and we analyze the dependance on the non minimal coupling to gravity.
The considered geometries have all the matter content isolated in a thin-shell and in this way we could analyze the influence of the coupling on the scalar field solutions and the consequences over the self-force.
Additionally, the obtained results clarify some aspects of the self-force in terms of the global properties of a given background geometry.
On the other hand, resonances of the field appear at critical values $\xi_c$ of the coupling which depend on the shape of the profile function of the metric or, more specifically, on the value of the deficit angles. We established that the stable domain of the coupling is related to the kind of matter to which it couples or, equivalently, to the sign of the trace of the extrinsic curvature tensor jump over the shell. 

We shall consider two models which are presented in Section \ref{s2}; in the first one, an interior and an exterior submanifolds are joined at an hypersurface where a thin-shell is placed and, in the second one, two exterior submanifolds are joined. The thin-shell is characterized by the trace of the jump on the extrinsic curvature tensor $\k$. Section \ref{s3} is divided in three subsections; in \ref{s3.1} and \ref{s3.2} the field equation sourced by the static scalar charge is solved in both types of manifolds by the method of separation of variables in cylindrical coordinates. The field $\P$ is split in two terms, one homogeneous and the other inhomogeneous at the position of the charge. Resonant configurations for which the field diverges for critical values $\xi_c$ of the coupling are found in Subsection \ref{s3.3}. 
These are associated to instabilities of the coupled scalar field equation in the corresponding background geometries. 
The scalar field is regularized at the position of the particle in Section \ref{s4} by subtracting the Detweiler-Whiting singular Green function to the actual field. Finally the self-force is calculated over the scalar charge evaluating the gradient of the regular field at the position of the particle. The results are analyzed analytically and numerically in terms of the coupling constant for different configurations of the background spacetime in \ref{s4.1} and \ref{s4.2}. Throughout the article the geometrized unit system is used where $c=G=1$.

\section{Approach}
\lb{s2}

The system is given by the action $S = S_{\P}+S_{m_0}+S_{q}$ in a fixed background.
The first term is the action for the free massless scalar field
\be
S_{\P} = -\frac{1}{ 8 \pi} \int \,d^{4}x \sqrt{-g} \lp g^{\a \b} \p_{\a}\P \,\p_{\b}\P + \xi R \, \P^{2} \rp \,,
\ee
where the integration is over all the spacetime with metric $g_{\a \b}$, $g$ is its determinant and $\xi$ the arbitrary coupling of the field to the curvature scalar $R$. The particle action $ S_{m_0} $ is
\be
S_{m_0} = - m_0 \int_{\gamma} d\tau \,,
\ee
where $\g$ is the scalar particle's world line, $m_0$ its bare mass and $d\tau = \sqrt{-g_{\a \b} \, \dot{z}^{\a} \dot{z}^{\b}} d\tau$ is the proper time differential along $\g$. 
The interaction term $ S_{q} $ coupling the field to the particle's charge $q$ is written as
\begin{eqnarray}
S_{q} &=& \int_{\gamma} d\tau \, q\,\Phi(z(\tau)) = q\int_{\g} d\tau  \int d^{4}x\sqrt{-g}\, \delta_{4}(x,z(\tau)) \,\Phi(x) \\
 &=&  q \,\int d^{4}x \sqrt{-g}\, \int_{\g} d\tau \,\delta_{4}(x,z(\tau)) \Phi(x)\,.
\end{eqnarray}

The wave equation for the coupled to gravity massless scalar field is obtained demanding the total action to be stationary under a variation $ \delta\Phi(x)$. 
We are interested in a scalar charge $q$ at rest; this yields the inhomogeneous equation
\be \lb{wave_eq}
\lp \square - \xi R \rp \P(x) = - 4 \pi q \frac{\d^3 (\bf x - \bf x')}{\dot{z}^{0} \,\sqrt{-g}} \,,
\ee
where $\square$ is the d'Alambertian operator of the metric, $\bf x' $ is the spatial position of the charge $q$, 
and  $\dot{z}^{0}  = dt/d\tau$
.
On the other hand, demanding the action to be stationary under a variation $\d z^{\a}(\tau)$ of the world line yields the equations of motion
\be
m(\tau) \frac{D \dot{z}^{\a}}{d \tau} 
= 
q \lp g^{\a\b} + \dot{z}^{\a} \dot{z}^{\b} \rp \nabla_{\b} \P(z)
\ee
for the scalar particle. The otherwise dynamical mass $m(\tau) = m_0 - q \P(z)$ is constant if the particle is at rest in a static spacetime \cite{ppv}. To hold the charge fixed, the total force exerted by a mechanical strut is 
\be
F_{strut}^{\a} = m(\tau) \G^{\a}_{0 0} \dot{z}^{0} \dot{z}^{0} - q \, g^{\a \b} \p_{\b} \P \,,
\ee
the second term corresponds to minus the scalar force over the static particle and will need to be regularized at the position of the charge, while the term with the Christoffel symbol $\G^{\a}_{00}$ is zero in locally flat geometries.
Specifically, $\G^{\a}_{00}=0$ is the case of the cylindrical thin-shell spacetimes we will consider which are described by line elements
\be \lb{metric}
ds^2 = - dt^2 + dr^2 + \rho^2(r) d\theta^2 + dz^2\,.
\ee
In an infinitely thin straight cosmic string geometry (without a shell), the profile function is simply $\rho(r) = r \, \om$ with the radial coordinate $0< r < \infty$, $0 \leq \theta < 2\pi$, and the parameter $ 0 < \om \leq 1$ which gives the deficit angle $2\pi (1-\om)$ of the conical manifold. For thin-shell spacetimes we will have two types of profile function. The first one is
\be
\mbox{Type I}: \; \rho(r) = \left\{ \ba{ll}
          \, r \, \om_i \,, & \mbox{if $0<r \leq r_i$ ($\mathcal{M}_{i}$)}\ \\
\lp r - r_i + r_e \rp \om_e \,, & \mbox{if $r_i \leq r < \infty$ ($\mathcal{M}_{e}$)}\,,
         \ea \right.
\ee
where the parameters $\om_i$ and $\om_e$ give the deficit angles in the interior $\mathcal{M}_{i}$ and exterior $\mathcal{M}_{e}$ regions, and the internal and external radii are related by the first fundamental form over the thin-shell hypersurface; $\r(r_i) = r_i  \om_i = r_e  \om_e$. The second type of profile corresponds to a wormhole spacetime with two asymptotic regions, this is
\be
\mbox{Type II}: \; \rho(r) = \left\{ \ba{ll}
        \lp r_- - r \rp \om_- \,, & \mbox{if $- \infty < r \leq 0$ ($\mathcal{M}_{-}$)}\ \\
        \lp r + r_+ \rp \om_+ \,, & \mbox{if $0 \leq r < + \infty$ ($\mathcal{M}_{+}$)}\,,
         \ea \right.
\ee
where two exterior regions $\mathcal{M}_{-}$ and $\mathcal{M}_{+}$, with $\om_-$ and $\om_+$ respectively, are joined over the hypersurface of a thin throat by the condition $\r(0) = r_- \om_ - = r_+ \om_+$. For example, a cylindrical wormhole which is symmetric across the throat with radius $r_0 = r_- = r_+$ has profile function $\rho(r) = \lp r_0 + |r|  \rp \om $.
Either Type I or Type II geometries are everywhere flat except at the shell hypersurface and, for Type I spacetimes only, in the case of conical interiors ($\om_i \neq 1$) the central axes represents a conical singularity. For both types, the Ricci scalar is $R(r)= -2 \, \k \, \d(r-r_s)$, where $r_s$ is the corresponding shell's radial position and $\k$ is the trace of the jump on the extrinsic curvature tensor over the thin-shell. In general: for Type I we have $\k = (\om_e - \om_i)/\r(r_s)$ with $r_s=r_i$, and for Type II $\k = (\om_+ + \om_-)/\r(r_s)$, with $r_s=0$.\\



\section{Scalar field in conical spacetimes}
\lb{s3}

The solution to Eq. (\ref{wave_eq}) in the conical manifold of an infinitely thin straight cosmic string without a thin-shell, $\rho(r) = r\,\om$, is known from 
\cite{fulling} and can be written in a close form as
\footnote{The axis at $r=0$ of the conical geometry, where the infinitely thin cosmic string would be placed, is intentionally excluded. The field in the manifold which includes the axis couples to the Ricci scalar $R(r) = - 2 (\om - 1) \delta(r)/(\om r)$ at $r=0$ and it can be shown to be $\P = \P_{\om} + \P_{\xi}$, with 
\be
\P_{\xi} = - \frac{q}{\pi \om} \frac{u}{rr' \sinh{u}}  \lim_{\e \to 0}\lbr \frac{(1-\om)\xi}{\om + 2 (1- \om)\xi K_0(\e)} \rbr 
\ee
where the term in brackets becomes identically null, \textit{i.e.} the complete solution which accounts for the coupling at the conical peak does not affect the relevant part of the field \cite{allen}.} 
 
\begin{flalign} \lb{cs}
\P_{\om} & = \frac{q}{\pi \sqrt{2 r r'}} \int \limits_{u}^{+\infty}\frac{ \sinh(\z/\om) \; \om^{-1} \quad d\z}{\left[ \cosh(\z/\om) - \cos{(\t- \t')} \right] \lp \cosh \z - \cosh u \rp^{1/2}}  & \\
&\mbox{where:}\; \cosh u = \frac{r^2 + r'^2 + (z-z')^2}{2 r r'} \,, \quad u  \geqslant 0 \,. & \nonumber
\end{flalign} 
For the static charge placed in Type I or Type II geometries, the field equation (\ref{wave_eq}) in cylindrical coordinates becomes
\be \lb{field_eq}  
\lbr \frac{\p^2}{\p r^2} + \frac{\r'(r)}{\r(r)} \frac{\p}{\p r} + \frac{1}{\r^2(r)} \frac{\p^2}{\p \t^2} + \frac{\p^2}{\p z^2 } - \xi\, R(r) \rbr \P = - 4\pi q \,\frac{\delta^{3}({\bf x} - {\bf x}')}{\r(r)}\,.
\ee
We look for a solution of the form
\be\lb{exp}
\P=q  \sum_{n=0}^{+\infty} \frac{4}{\pi \lp 1+\d_{0,n} \rp} \, Q_n(\t) \int \limits_{0}^{+\infty}dk\,Z(k,z)\chi_{n}(k,r)  \, ,
\ee
where $F_z=\{Z(k,z)=\cos[k(z-z')]\}$ and $F_{\t}=\{ Q_n(\t)=  \cos[n(\t-\t')]\}$ are a complete set of orthogonal functions of the coordinates $z$ and $\t$. Then, the radial functions $\chi_{n}(k,r)$ are obtained from the radial equation
\be \lb{radial_eq}
\lk \frac{\p}{\p r} \lbr \r(r) \frac{\p}{\p r} \rbr - \r(r)\lbr \lp \frac{n}{\r(r)} \rp^2 + k^2 + \xi\, R(r) \rbr \rk \chi_n(k,r) =  - \delta(r - r') \,.
\ee
Integrating over an infinitesimal radial interval around the position of the shell $r=r_s$ on (\ref{radial_eq}) we obtain the condition
\be \lb{jump_rs}
\lbr \frac{\p}{\p r} \, \chi_n(k,r) \rbr^{r={r_s}^+}_{r={r_s}^-} = - 2\xi\, \k\, \chi_n(k,r)\big|_{r=r_s} 
\ee
for the radial solutions, where we used the Ricci scalar $R(r)= -2 \k \d(r-r_s)$ and the continuity of the profile function $\r(r)$ and of $\chi_n(k,r)$ at $r=r_s$. Analogously, from (\ref{radial_eq}),
\be \lb{jump_r'}
\lbr \frac{\p}{\p r} \, \chi_n(k,r) \rbr^{r={r'}^+}_{r={r'}^-} = - \frac{1}{\r(r')} \, ,
\ee
assuming the continuity of $\chi_n(k,r)$ at the radial position $r=r'$ of the particle. The radial solutions in Type I and Type II geometries will be obtained from (\ref{radial_eq}) plus the latter conditions and the requirements over the corresponding asymptotic regions. In a conical manifold without shell, 
(\ref{radial_eq}) reduces to
the $\textit{n}^{th}$ order inhomogeneous Bessel equation
\be\lb{bessel}
\left[\frac{\p^2}{\p r^2}+\frac{1}{r}\frac{\p}{\p r}-\left(k^2+\frac{\upsilon^2}{r^2}\right)\right] \chi^{\om}
=
-\frac{\d(r-r')}{\om r}\,,
\ee
where $\upsilon=n/\om$, for each $n$ and $k$. Requiring finiteness if $r\rightarrow 0$ and $r\rightarrow\infty$, the continuity at $r=r'$ and its derivative discontinuity (\ref{jump_r'}), we obtain the scalar field in the infinitely thin straight cosmic string spacetime given as a series expansion with the radial solutions:
\be \lb{csrf}
\chi^{\om}
= \frac{1}{\om} K_{\upsilon}(kr_>)I_{\upsilon}(kr_<) \,, 
\ee
where $K_{\upsilon}(kr)$ and $I_{\upsilon}(kr)$ are the usual modified Bessel functions of order $\upsilon=n/\om$ with $n \, \epsilon \, \mathbb{N}_0$, two independent solutions of the homogeneous version of Eq. (\ref{bessel}), while 
$r_< = \mbox{min}\left\{ r ; r' \right\}$ and $r_> = \mbox{max}\left\{ r ; r' \right\}$. The field $\P_{\om}$ given as the series expansion (\ref{exp}) with $\chi^{\om}$ is equivalent to the integral form (\ref{cs}). \\


\subsection{Scalar field in Type I spacetimes}
\lb{s3.1}

The radial solutions in Type I spacetimes will be obtained defining an interior radial coordinate $r_1= r$ for $0<r<r_i$, in the interior region $\mathcal{M}_{i}$ of parameter $\om_i$, and an exterior radial coordinate $r_2 =  r - r_i + r_e$ for $r_i < r < \infty$, in the exterior region $\mathcal{M}_{e}$ with parameter $\om_e$.
Using these new coordinates we write the radial functions as
\be \lb{radial_I}
\chi_n(k,r)= \lk \ba{ll} 
\ba{ll} 
\chi^i(r_1)  \quad &\mbox{for} \; \mathcal{M}_{i}  \\
\chi^e(r_2) \quad &\mbox{for} \; \mathcal{M}_{e}
\ea 
\ea \right.
\ee
for each $n$ and $k$. From (\ref{radial_eq}), the respective equations become
\be \lb{radial_eq_ie}
\ba{ll} 
\left[\frac{\p^2}{\p r_1^2} + \frac{1}{r_1}\frac{\p}{\p r_1} - \left(k^2 + \frac{\l^2}{r_1^2}\right) \right] \chi^{i}
&= \lk \ba{ll} 
- \frac{\d(r_1 - r'_1)}{\om_i r_i}\,, & \mbox{x' in}\,\mathcal{M}_{i} \\
0 \,, & \mbox{x' in}\,\mathcal{M}_{e}
\ea \right. \\
\left[\frac{\p^2}{\p r_2^2} + \frac{1}{r_2}\frac{\p}{\p r_2} - \left(k^2 + \frac{\n^2}{r_2^2}\right) \right] \chi^{e}
&= \lk \ba{ll} 
0 \,,  & \mbox{x' in}\,\mathcal{M}_{i}
 \\
- \frac{\d(r_2 - r'_2)}{\om_e r_e}  \,, &\mbox{x' in}\,\mathcal{M}_{e}
\ea \right.
\ea
\ee
with $\l = n/\om_i$ and $\n = n/\om_e $. To solve them we impose: the boundary conditions at the axis and infinity
\be \lb{0_inf}
\lim_{r_1 \rightarrow 0} \chi^{i} \neq \infty \,, \quad
\lim_{r_2 \rightarrow +\infty} \chi^{e} = 0 \,,
\ee
continuity of the field and its derivative discontinuity (\ref{jump_rs}) at the shell
\be \lb{cont_rie}
\chi^{i}(r_1=r_i) = \chi^{e}(r_2=r_e) \,,
\ee
\be \lb{jump_rie}
\frac{\p}{\p r_2} \chi^{e}\Bigl|_{r_2=r_e} - \frac{\p}{\p r_1} \chi^{i} \Bigl|_{r_1=r_i} = - 2 \k \xi \,\chi^{i}(r_i) \,,
\ee
plus the conditions for the inhomogeneity at the position of the charge. Considering first the particle placed in $\mathcal{M}_{i}$, the inhomogeneity discontinuity (\ref{jump_r'}) at $r_1' = r'$ is
\be \lb{jump_i}
 \lp \frac{\p}{\p r_1} \chi^{i}\rp \biggl|^{{r'_1}^{+}}_{{r'_1}^{-}} = - \frac{1}{\om_i r'_1} \,,
\ee
while $\chi^{i}$ is required to be continuous at $r_1=r_1'$. From (\ref{radial_eq_ie}) and (\ref{0_inf}), with the charge in $\mathcal{M}_{i}$, the radial solutions are given by
\begin{flalign} 
\chi^{i} = & \, \chi^{\om_i} + \frac{1}{\om_i} I_{\l}(k r_1)  A_n(k) \;\;,\; \l = n/\om_i \,; \lb{radial_int}&\\
\chi^{e}  = &\, \frac{1}{\om_i} K_{\n}(k r_2) C_n(k) \;\;,\; \n = n/\om_e \,;  \lb{radial_e}&
\end{flalign}
where the terms
\begin{flalign}
\chi^{\om_i} = & \, \frac{1}{\om_i} I_{\l}(k r_{{1}_<}) K_{\l}(k r_{{1}_>})  \,,  \; \ba{ll} 
r_{{1}_<} = \mbox{min}\{ r_1, r_1' \} \\
r_{{1}_>} = \mbox{max}\{ r_1, r_1' \}  
\ea &
\end{flalign}
account for the derivative's jump (\ref{jump_i}) of $\chi^{i}$. The terms $\chi^{\om_i}$ of the interior solution are intentionally left apart because, inside the Fourier series, they sum up the integral expression (\ref{cs}), as in (\ref{csrf}). Finally, the coefficients $A_n(k)$ and $C_n(k)$ are determined from conditions (\ref{cont_rie}) and (\ref{jump_rie}) at the shell;
\begin{flalign} \lb{A_n}
A_{n}(k) = & - I_{\l}(kr'_1)  \times &\nonumber \\
&\frac{K_{\l}(k r_i) K'_{\n}(k r_e) - K_{\n}(k r_e) K'_{\l}(kr_i) + 2 \k \xi K_{\n}(k r_e) K_{\l}(k r_i)}{I_{\l}(k r_i) K'_{\n}(k r_e) - K_{\n}(k r_e) I'_{\l}(k r_i) + 2 \k \xi K_{\n}(k r_e) I_{\l}(k r_i)}\,,
&
\end{flalign}
\begin{flalign} \lb{C_n}
C_n(k) = &
\frac{A_{n}(k) I_{\l}(kr_i) + I_{\l}(kr'_1)K_{\l}(kr_i)}{K_{\n}(k r_e)} \,,&
\end{flalign}
with the prime over Bessel functions implying derivative with respect to the corresponding radial coordinate.
Alternatively, with the particle in $\mathcal{M}_{e}$ the inhomogeneity is at $r_2'= r' - r_i + r_e$ and now
\be \lb{jump_e}
 \lp \frac{\p}{\p r_2} \chi^{e}\rp \biggl|^{{r'_2}^{+}}_{{r'_2}^{-}} = - \frac{1}{\om_e r'_2} \, , 
\ee
while $\chi^{e}$ is required to be continuous at $r_2=r_2'$. From (\ref{radial_eq_ie}) and (\ref{0_inf}), with the charge in $\mathcal{M}_{e}$, the radial solutions are given by
\begin{flalign}
\chi^{i} = & \frac{1}{\om_e} I_{\l}(k r_1)  D_n(k) \;\;,\; \l = n/\om_i \,; \lb{radial_i} &\\
\chi^{e}  = &\, \chi^{\om_e} + \frac{1}{\om_e} K_{\n}(k r_2) B_n(k) \;\;,\; \n = n/\om_e \,; \lb{radial_ext} & 
\end{flalign}
where the terms
\begin{flalign}
\chi^{\om_e} = & \, \frac{1}{\om_e} I_{\n}(k r_{{2}_<}) K_{\n}(k r_{{2}_>})  \,,  \; \ba{ll} 
r_{{2}_<} = \mbox{min}\{ r_2, r_2' \} \\
r_{{2}_>} = \mbox{max}\{ r_2, r_2' \}  
\ea &
\end{flalign}
account for the derivative's jump (\ref{jump_e}) of $\chi^{e}$, and the coefficients $B_n(k)$, $D_n(k)$ are determined from (\ref{cont_rie}) and (\ref{jump_rie}) at the shell;
\begin{flalign} \lb{B_n}
B_{n}(k) = & - K_{\n}(kr'_2)  \times &\\
& \frac{I_{\l}(k r_i) I'_{\n}(k r_e) - I_{\n}(k r_e) I'_{\l}(kr_i) + 2 \k \xi I_{\n}(k r_e) I_{\l}(k r_i)}{I_{\l}(k r_i) K'_{\n}(k r_e) - K_{\n}(k r_e) I'_{\l}(k r_i) + 2 \k \xi K_{\n}(k r_e) I_{\l}(k r_i)} \,,&\nonumber 
\end{flalign}
\begin{flalign} \lb{D_n}
D_n(k) = &
\frac{I_{\n}(k r_e) K_{\n}(k r'_2) + B_{n}(k) K_{\n}(k r_e)}{I_{\l}(kr_i)} \,.&
\end{flalign}

Either with the particle in the interior or exterior region, we see from the two terms explicitly written in the respective radial solutions, (\ref{radial_int}) or (\ref{radial_ext}), that in the four-dimensional region where the charge is placed we can separate the field in two parts as
\be \lb{split_I}
\Phi = \P_{\om} + \P_{\xi}\, \quad \mbox{if $(x, x') \in \mathcal{M}_i$ or $\mathcal{M}_e$}\,.
\ee
The field $\P_{\om}$, corresponding to the series with radial functions called $\chi^{\om}$, is inhomogeneous at $x'$ and locally equivalent to the field (\ref{cs}) of a charge in a cosmic string manifold with the corresponding parameter $\om=\om_{i}$ or $\om=\om_{e}$. The remaining part of the field, the series called $\P_{\xi}$, is homogeneous at $x'$ and accounts for the distortion produced in the specific thin-shell geometry with information about the coupling and boundary conditions.


\subsection{Scalar field in Type II thin-shell wormholes}
\lb{s3.2}

The radial solutions in Type II geometries, corresponding to wormholes with two asymptotic regions, will be obtained defining the radial coordinate $r_1(r) = r_- - r$  for $r<0$, in the region $\mathcal{M}_{-}$ of parameter $\om_-$, and the radial coordinate $r_2(r) = r + r_+$ for $r > 0$, in the region $\mathcal{M}_{+}$ of parameter $\om_+$. 
Using these coordinates we write the radial functions as
\be \lb{radial_II}
\chi_n(k,r)= \lk \ba{ll} 
\ba{ll} 
\chi^-(r_1)  \quad &\mbox{for} \; \mathcal{M}_{-}  \\
\chi^+(r_2) \quad &\mbox{for} \; \mathcal{M}_{+}
\ea 
\ea \right.
\ee
for each $n$ and $k$. 
With the particle placed in $\mathcal{M}_{+}$, the inhomogeneity is at $r_2' > r_+$, the radial equations are
\be \lb{radial_eq_wh}
\ba{ll} 
\left[\frac{\p^2}{\p r_1^2} + \frac{1}{r_1}\frac{\p}{\p r_1} - \left(k^2 + \frac{\l^2}{r_1^2}\right) \right] \chi^{-}
= 
&0 
\\
\left[\frac{\p^2}{\p r_2^2} + \frac{1}{r_2}\frac{\p}{\p r_2} - \left(k^2 + \frac{\n^2}{r_2^2}\right) \right] \chi^{+}
= 
&- \frac{\d(r_2 - r'_2)}{\om_+ r_+}  
\ea
\ee
with $\l = n/\om_-$ and $\n = n/\om_+$. The boundary conditions at the infinities are
\be \lb{inf_-+}
\lim_{r_1 \rightarrow +\infty} \chi^{-} = 0 \,, \qquad
\lim_{r_2 \rightarrow +\infty} \chi^{+} = 0 \,,
\ee
continuity of the field and its derivative discontinuity (\ref{jump_rs}) at the shell are
\be \lb{cont_wh}
\chi^{-}(r_1=r_-) = \chi^{+}(r_2=r_+)\,,
\ee
\be \lb{jump_wh}
\frac{\p}{\p r_2} \chi^{+}\Bigl|_{r_2=r_+} + \frac{\p}{\p r_1} \chi^{-} \Bigl|_{r_1=r_-} = - 2 \k \xi \,\chi^{-}(r_-) \,,
\ee
respectively, and the inhomogeneity discontinuity (\ref{jump_r'}) is
\be \lb{jump_+}
 \lp \frac{\p}{\p r_2} \chi^{+}\rp \biggl|^{{r'_2}^{+}}_{{r'_2}^{-}} = - \frac{1}{\om_+ r'_2} \, ,
\ee
while $\chi^{+}$ is required to be continuous at $r_2=r_2'$. From (\ref{radial_eq_wh}) and (\ref{inf_-+}) the radial solutions are given by
\begin{flalign}
\chi^{-} = &\frac{1}{\om_+} K_{\l}(k r_1)  E_n(k) \;\;,\; \l = n/\om_- \,; \lb{radial_wh_-}&\\
\chi^{+}  = & \, \chi^{\om_+} + \, \frac{1}{\om_+} K_{\n}(k r_2) W_n(k) \;\;,\; \n = n/\om_+ \,; \lb{radial_wh_+}&
\end{flalign}
where the terms
\begin{flalign}
\chi^{\om_+} = & \, \frac{1}{\om_+} I_{\n}(k r_{{2}_<}) K_{\n}(k r_{{2}_>})  \,,  \; \ba{ll} 
r_{{1}_<} = \mbox{min}\{ r_2, r_2' \} \\
r_{{1}_>} = \mbox{max}\{ r_2, r_2' \}  
\ea &
\end{flalign}
account for the derivative's jump (\ref{jump_+}) of $\chi^{+}$. The coefficients $E_n(k)$ and $W_n(k)$ are determined from (\ref{cont_wh}) and (\ref{jump_wh}) at the shell;
\begin{flalign} \lb{W_n}
W_n(k) = & -  K_\n(k r'_2) \times & \\
&\frac{I_{\n}(k r_+) K'_{\l}(kr_-) + K_{\l}(k r_-)I'_{\n}(k r_+) + 2 \k \xi I_{\n}(k r_+) K_{\l}(k r_-)}{K_{\l}(k r_-) K'_{\n}(k r_+) + K_{\n}(k r_+) K'_{\l}(k r_-) + 2 \k \xi K_{\n}(k r_+) K_{\l}(k r_-)} \,,& \nonumber
\end{flalign}
\begin{flalign} \lb{E_n}
E_n(k) =  &
\frac{I_{\n}(k r_+)K_{\n}(k r'_2) + W_{n}(k) K_{\n}(k r_+)}{K_{\l}(kr_-)} \,,&
\end{flalign}
with the prime over Bessel functions implying derivative with respect to the corresponding radial coordinate. As we pointed out earlier in Subsection \ref{s3.1} for the solutions in Type I geometries, in the four-dimensional region where the charge is placed, we can separate the field as
\be\lb{split_II}
\Phi = \P_{\om_+} + \P_{\xi}\, \quad \mbox{if $(x, x') \in \mathcal{M}_+$}\,,
\ee
with $\P_{\om_+}$ constructed with the radial solutions $\chi^{\om_+}$, inhomogeneous at $x'$ and locally equivalent to the field (\ref{cs}) of a charge in a cosmic string manifold, and the term $\P_{\xi}$ which is homogeneous at $x'$.


\subsection{Resonant configurations}
\lb{s3.3}
The scalar field produced by a static scalar particle of charge $q$ in Type I or Type II spacetimes is 
\be\lb{scalar_field}
\P =  \frac{4q}{\pi} \sum_{n=0}^{+\infty} \frac{\cos[n(\t-\t')]}{\lp 1+\d_{0,n} \rp} \,  \int \limits_{0}^{+\infty}dk\, \chi_{n}(k,r) \cos[k(z-z')]   \, ,
\ee
with the radial functions $\chi_{n}(k,r)$ specified in the previous subsections. The dependence of the field on the curvature coupling is seen from the respective coefficients (\ref{A_n})-(\ref{C_n}), (\ref{B_n})-(\ref{D_n}) or (\ref{W_n})-(\ref{E_n}) in each of the radial solutions. 
From a simple inspection on the results obtained in Type I spacetimes, we see that the denominators of coefficients $A_n(k)$ and $B_n(k)$, given in (\ref{A_n}) and (\ref{B_n}) respectively, do not vanish if $\k$ is negative and the coupling $\xi$ is positive. The same is guaranteed with a positive $\k$ and a negative value for $\xi$, including denominators of coefficients $W_n(k)$ obtained for Type II, given in (\ref{W_n}), where $\k$ is always positive.
In the contrary, if the product $\k \, \xi$ is positive the coefficients manifest the possibility of encountering poles in the integrands of (\ref{scalar_field}) for some value $k=k_p$.
Nevertheless, if the integrand of a given mode $n$ presents a pole at $k_p > 0$, the divergency can be circumvented splitting the integral at $k_p \mp \e$, with $\e \to 0$, by canceling out the contributions of the lateral limits. 
As an alternative to the last method, contour integrals for the complex-valued integrand function can be used to obtain the positive real half-line integral.
But if the denominator in the integrand coefficient is null for $k=0$ we have, inevitably, a divergent mode. 
We can examine directly the integrand of each mode in the limit $k \to 0^+$ to identify these divergencies. For example, in Type I geometries with the particle placed at $r_1=r_1'$, we obtained the internal radial solutions $\chi^{i}_n(k,r_1) = \, \chi^{\om_i}_n(k,r_1) + \frac{1}{\om_i} I_{\l}(k r_1)  A_n(k)$, and the external ones $\chi^{e}_n(k,r_2) = \frac{1}{\om_i} K_{\n}(k r_2)  C_n(k)$, given in (\ref{radial_int}) and (\ref{radial_e}), from which we can compute:
\begin{flalign} 
\chi^{i}_n(k,r_1)   \xrightarrow{k \to 0^+} & \;
\frac{ 1 }{2 n } \lbr \lp \frac{r_{1_<}}{ r_{1_>}}\rp^{n/\om_i}
- \frac{ \xi }{\lp \xi - \xi_c^{(n)} \rp} \lp \frac{r_1 \, r_1'}{{r_i}^2}\rp^{n/\om_i} \rbr& \nonumber\\
&= 
\frac{ 1 }{2 n } \lbr 1 - \frac{ \xi }{\lp \xi - \xi_c^{(n)} \rp} \lp \frac{r_{1_>}}{{r_i}}\rp^{2n/\om_i} \rbr  \lp \frac{r_{1_<}}{ r_{1_>}}\rp^{n/\om_i}\,,
&
\end{flalign}
\begin{flalign} 
\chi^{e}_n(k,r_2)   \xrightarrow{k \to 0^+} & \;
- \frac{ 1 }{2 \,\lp \xi - \xi_c^{(n)} \rp (\om_e - \om_i)} \lp \frac{r_1'}{{r_i}}\rp^{n/\om_i} \lp \frac{r_e}{{r_2}}\rp^{n/\om_e} \,. &
\end{flalign}
The critical value $\xi_c^{(n)}$ is given by
\be
\xi_c^{(n)} = \frac{n}{ \r(r_i) \, \k } = \frac{n}{\om_e-\om_i} \,,
\ee
for $n \in \mathbb{N}$, in Type I spacetimes. Equivalently, with the particle placed at $r_2 = r_2'$ we had the internal and external radial solutions: $\chi^{i}_n(k,r_1)  =\frac{1}{\om_e} I_{\l}(k r_1) D_n(k)$ and $\chi^{e}_n(k,r_2)  = \chi^{\om_e}_n(k,r_2) + \frac{1}{\om_e} K_{\n}(k r_2) B_n(k)$, given in (\ref{radial_i}) and (\ref{radial_ext}), from where we can see analogously
\begin{flalign} 
\chi^{i}_n(k,r_1) \xrightarrow{k \to 0^+} 
&
- \frac{ 1 }{2 \,\lp \xi - \xi_c^{(n)} \rp (\om_e - \om_i)} \lp \frac{r_1}{{r_i}}\rp^{n/\om_i} \lp \frac{r_e}{{r_2'}}\rp^{n/\om_e} \,,&
\end{flalign}
\begin{flalign} 
\; \chi^{e}_n(k,r_2)   \xrightarrow{k \to 0^+} & \;
\frac{ 1 }{2 n } \lbr \lp \frac{r_{2_<}}{ r_{2_>}}\rp^{n/\om_e}
- \frac{ \xi }{\lp \xi - \xi_c^{(n)} \rp} \lp \frac{r_2 \, r_2'}{{r_e}^2}\rp^{-n/\om_e} \rbr &\nonumber
  \\
&= 
\frac{ 1 }{2 n } \lbr 1 - \frac{ \xi }{\lp \xi - \xi_c^{(n)} \rp} \lp \frac{r_e}{r_{2_<}}\rp^{2n/\om_e} \rbr  \lp \frac{r_{2_<}}{ r_{2_>}}\rp^{n/\om_e} \,.
&
\end{flalign}
These limits show a divergence for a mode $n \geqslant 1$ if the curvature coupling takes the critical value $\xi_c^{(n)} = n/(\om_e-\om_i)$, which is identified with a resonant configuration for the $n^{th}$ mode of the scalar field in Type I spacetimes. For $n=0$ there is no value of $\xi$ associated to this kind of resonances (the denominator of radial coefficients do not vanish for $k=0$ in the $n=0$ mode). 
In terms of the extrinsic curvature on the thin-shell in Type I spacetimes; if $\k < 0$ (ordinary matter thin-shell) the coupling can take values $\xi > \xi_c^{(n=1)} 
= 1/(\k \om_i r_i)$ to avoid encountering a resonant mode which makes the field divergent and, on the other hand, if $\k >0$ (exotic matter thin-shell) couplings $\xi < \xi_c^{(n=1)} = 1/(\k \om_i r_i)$ ensure that the configuration does not produce some resonant mode.
Repeating the analysis in Type II geometries, with the radial solutions obtained with the particle placed at $r_2=r_2'$: $\chi^{-}_n(k,r_1) = \frac{1}{\om_+} K_{\l}(k r_1)  E_n(k)$ for region $\mathcal{M}_-$, and $\chi^{+}_n(k,r_2) = \, \chi^{\om_+}_n(k,r_2) + \frac{1}{\om_+} K_{\n}(k r_2)  W_n(k)$ for region $\mathcal{M}_+$, given in (\ref{radial_wh_-}) and (\ref{radial_wh_+}), we can see the same dependance
\begin{flalign} 
\chi^{-}_n(k,r_1) \xrightarrow{k \to 0^+} & \;
- \frac{ 1 }{2 \,\lp \xi - \xi_c^{(n)} \rp (\om_+ + \om_-)} \lp \frac{r_-}{{r_1}}\rp^{n/\om_-} \lp \frac{r_+}{{r_2'}}\rp^{n/\om_+} \,,&
\end{flalign}
\begin{flalign} 
\chi^{+}_n(k,r_2)   \xrightarrow{k \to 0^+} & \;\frac{ 1 }{2 n } \lbr \lp \frac{r_{2_<}}{ r_{2_>}}\rp^{n/\om_+}
- \frac{ \xi }{\lp \xi - \xi_c^{(n)} \rp} \lp \frac{r_2 \, r_2'}{{r_+}^2}\rp^{-n/\om_+} \rbr& \nonumber\\
&= 
\frac{ 1 }{2 n } \lbr 1 - \frac{ \xi }{ \lp \xi - \xi_c^{(n)} \rp} \lp \frac{r_+}{r_{2_<}}\rp^{2n/\om_+} \rbr  \lp \frac{r_{2_<}}{ r_{2_>}}\rp^{n/\om_+} \,,
&
\end{flalign}
with 
\be
\xi_c^{(n)} = \frac{n}{ \r(0) \, \k} = \frac{n}{\om_+ +\om_-}
\ee
 for $n \in \mathbb{N}$, in the wormhole spacetime. 
These limits show a resonant mode $n \geqslant 1$ for the scalar field if the coupling takes the critical value $\xi_c^{(n)} = n/(\om_+ +\om_-)$.
To avoid encountering a resonant mode, which makes the field divergent in Type II spacetimes, the coupling must take values $\xi < \xi_c^{(n=1)}$.


\section{Regular field and scalar self-force}
\lb{s4}


The static scalar self-force over a charged particle in a curved spacetime is obtained regularizing the actual field with a singular field $\P^S$ as
\be \lb{force}
f_{\a} = q \lim_{\bf x \to \bf x'} \nabla_{\a} \lp \P - \P^S \rp\,,
\ee
where the coincidence limit takes the coordinate spatial components of the point $x$ to the charge's position $x'$ along the shortest geodesic connecting them \cite{casals}. The field $\P^S = q \, 4 \pi \, G_{DW}(x ; x')$ for a static particle in a static spacetime is constructed with the Detweiler-Whiting singular Green function in three dimensions over the normal convex neighborhood of $\bf x'$ \cite{DW}. This Green function has the same singularity structure as the particle's actual field, exerts no force on the particle and, in the considered static problem, 
can be calculated as \cite{ppv}
\be \lb{G_{DW}} 
G_{DW}(x ; x') = \frac{1}{4 \pi} \lp\frac{{\bf \bigtriangleup}^{1/2}}{\sqrt{2\s}} + \frac{1}{2} \int_{\tau_{ret}}^{\tau_{adv}} V(x, x'(\tau)) d\tau \rp \,,
\ee
where $\s=\s( \bf x, x')$ is half the squared geodesic distance between $\bf {x}$ and $\bf {x}'$ as measured in the purely spatial sections of the spacetime, the Van-Vleck Morette determinant has the following expansion 
\be
{\bf \bigtriangleup}^{1/2} = 1 + \frac{1}{12}R_{\a \b} \s^{; \a} \s^{; \b} + \mathcal{O}\left(\s^{3/2}\right) \,,
\ee
while the tail part is given by
\be
\int_{\tau_{ret}}^{\tau_{adv}} V(x, x'(\tau)) d\tau = \sqrt{2 \s} \lp \xi -  \frac{1}{6} \rp R  + \mathcal{O} \lp \s \rp \,.
\ee
The terms of order $\mathcal{O}\left(\s\right)/\sqrt{2\s}$ would be irrelevant for the renormalization of the field since they vanish in the coincidence limit. To regularize the field of a point scalar charge in conical spacetimes we will need the regular part of the inhomogeneous field in the locally identical geometry of a infinitely thin cosmic string manifold. Metric (\ref{metric}) is locally flat, except over the thin-shell, and an expression for the singular field can be simply given by the Green function over Minkowski spacetime, i.e. $\P^S = q /\lp \sqrt{2\s}\rp$. Then, this singular field can be represented as 
\begin{flalign} \lb{sing}
\P^S   \equiv \P_{Mink} &= \P_{\om =1} & \nonumber\\ 
&= \frac{q}{\pi \sqrt{2 r r'}} \int \limits_{u}^{+\infty}\frac{ \sinh(\z) \quad d\z}{\left[ \cosh(\z) - \cos{\lp \t- \t' \rp} \right] \lp \cosh \z - \cosh u \rp^{1/2}}  \,,& 
\end{flalign}
where for $\P_{\om=1}$ we used the integral expression (\ref{cs}) with $\om=1$. 

We first compute the self-force over a scalar charge in the infinitely thin straight cosmic string background, which only has a radial component due to cylindrical symmetry and
is equal to
\be
f_{\om}
= - \, q^2 \, \frac{L_{\om}}{4 \pi} \frac{1}{r'^2}
\ee
with
\be 
L_{\om}=\int \limits_{0}^{+\infty} \left[ \frac{\sinh(\zeta/\om)}{\om\left[\cosh(\zeta/\om)-1\right]}-\frac{\sinh{\zeta}}{\cosh{\zeta}-1}\right]\frac{d\zeta}
{\sinh(\zeta/2)} \,,
\ee
which was obtained evaluating $\t=\t'$ and $z=z'$, in (\ref{cs}) for the actual field and in (\ref{sing}) for the singular field, to take the coincidence limit (\ref{force}) over radial geodesics. The self-energy is defined as $U = \frac{q}{2} \, \P_{\om}^{ren}$, with the renormalized field obtained from 
\be \lb{cs_ren}
\P_{\om}^{ren} = \lim_{\bf x \to \bf x'} \lp \P_{\om} - \P^S \rp  =  \frac{q}{2 \pi} \frac{L_{\om}}{r'}  \,.
\ee
This self-field can be represented as a regular field at the position of the particle given by
\be \lb{reg_cs}
\P_{\om}^R(r,r') = \frac{q}{4\pi} \frac{L_{\om}}{r'} \lp 2 - \ln{\frac{r}{r'}} \rp \,, 
\ee
which is an homogeneous solution of the wave equation (\ref{wave_eq}), where the coincidence limits over the irrelevant coordinates are assumed in advance. The renormalized field (\ref{cs_ren}) is restored from the last result putting $r=r'$, while the self-force is $f_{\om} = q \, \lp \partial_r \P_{\om}^R \rp |_{r'} $. We note that the self-energy of a scalar particle in a conical manifold coincides with that of an electric charge, known as Linet's result on a cosmic string geometry \cite{linet}.\\

The singular structure in (\ref{sing}) is encountered in every conical geometry and the regularization procedure will be analogous in each of the conical thin-shell spacetimes because they are locally indistinguishable.
In virtue of the explicit splitting of the field as $\P = \P_{\om} + \P_{\xi}$, done in Subsection \ref{s3.1} for Type I (\ref{split_I}), or (\ref{split_II}) in \ref{s3.2} for Type II geometries,
the singular part at the position of the charge appears in the term $\P_{\om}$ and the regularization procedure yields 
\be \lb{regular}
\P^R = \P_{\om}^R + \P_{\xi} \,,
\ee
with the regular field $\P_{\om}^R$ given in (\ref{reg_cs}), with the corresponding parameter $\om$.

\subsection{Self-force in Type I geometries}
\lb{s4.1}

\begin{figure} [h!] 
\centering
\subfloat[vicinities of the shell.]
{\label{f1a}\includegraphics[width=7.5cm]{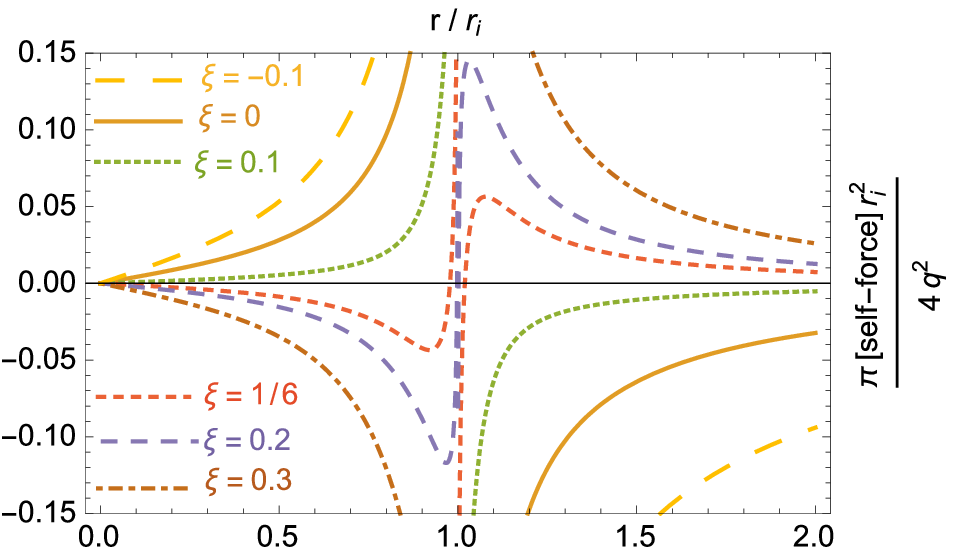}}
\subfloat[far from the shell.]
  {\label{f1b}
    \includegraphics[width=6.7cm]{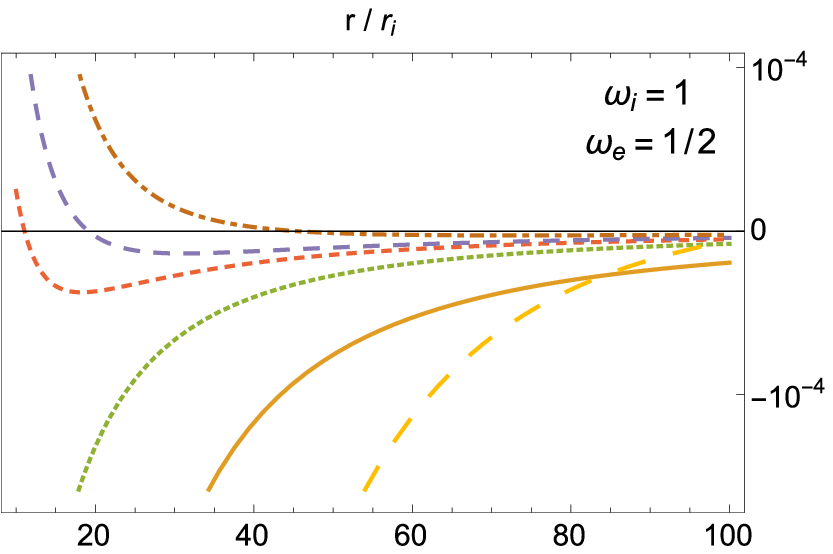}}
    \caption{Dimensionless self-force $\frac{\pi r_i^2}{4q^2} f$ as a function of $r/r_i$ in a Type I spacetime with a thin-shell of ordinary matter. In the conical exterior, the force is asymptotically attractive for any value of the coupling.
    } 
      \label{f1}
    \end{figure}

\begin{figure} [h!] 
\centering
\subfloat[vicinities of the shell.]
{\label{f2a}\includegraphics[width=7.5cm]{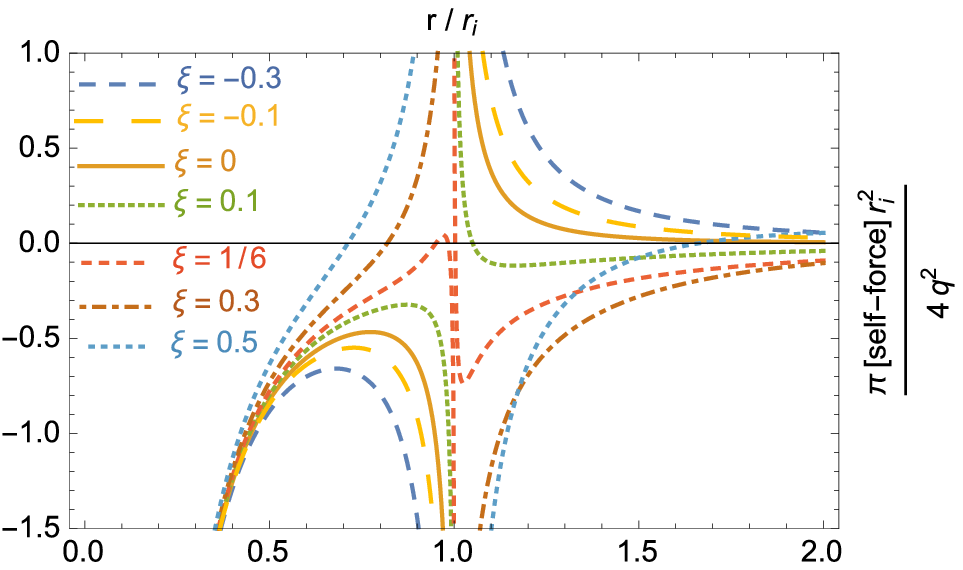}}
\subfloat[far from the shell.]
  {\label{f2b}
    \includegraphics[width=6.7cm]{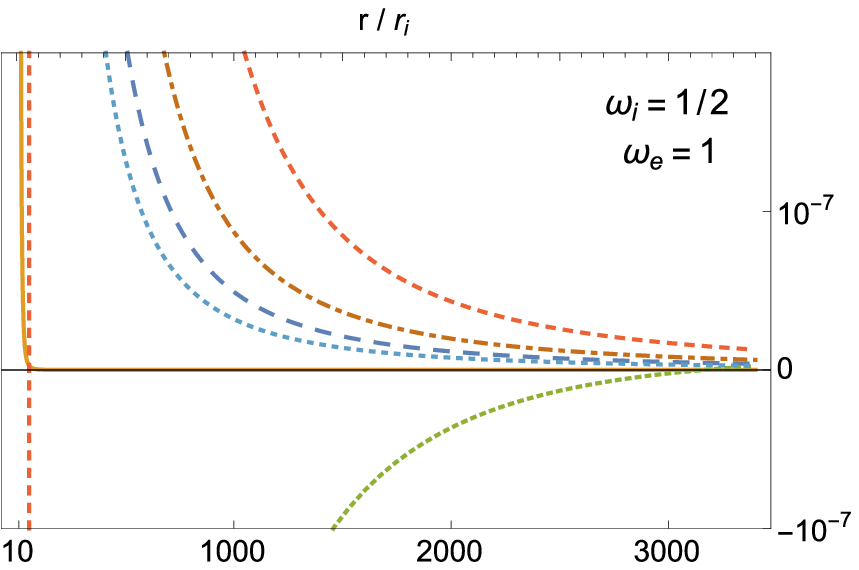}}
    \caption{Dimensionless self-force $\frac{\pi r_i^2}{4q^2} f$ as a function of $r/r_i$ in a Type I spacetime with a thin-shell of exotic matter. In a Minkowski exterior, the force is asymptotically repulsive for any value of the coupling.
    } 
      \label{f2}
    \end{figure}

The regularization procedure applied to the field obtained in Type I thin-shell geometries, with radial functions (\ref{radial_I}) given in terms of coordinates $r_1(r) = r$ for $\mathcal{M}_i$ or $r_2(r) = r - r_i + r_e$ for $\mathcal{M}_e$, yields the regular field at a general radial position $r'$ of the particle:
\be\lb{ren}
\P^{R}(r,r')=\left\{ \ba{ll}
 \frac{q\,L_{\om_i}}{4\pi r'}  \lp 2 - \ln{\frac{r}{r'}} \rp + 
\frac{4q}{\pi w_i} \sum\limits_{n=0} ^{+\infty} \int \limits_{0}^{+\infty}dk  \,  I_{\l}(k r) \frac{A_{n}(k)}{1+\delta_{n,0}}  \quad &\mbox{if $r' < r_i$ 
}\\
 \frac{q\, L_{\om_e}}{4\pi r'_2} \lp 2 - \ln{\frac{r_2(r)}{r'_2}} \rp + 
\frac{4q}{\pi w_e}  \sum\limits_{n=0}^{+\infty} \int \limits_{0}^{+\infty} dk \, K_{\n}(k r_2(r)) \frac{B_{n}(k)}{1+\delta_{n,0}}  \quad &\mbox{if $r' > r_i$ 
}
\ea\right.
\ee
from where the self-energy is obtained as $U_{self}=\frac{q}{2} \P^R|_{r=r'}$. If the charge is placed in a region without deficit angle ($\om = 1$), then $L_{\om} = 0$ and the first term is null. The radial scalar self-force $f= q \, \lp \partial_r \P^R \rp |_{r'}$ over the particle is 
\be \lb{f_I}
f=
\left\{\ba{ll}
- \frac{q^2 \, L_{\om_i}}{4\pi \, r'^2}  + 
\frac{4q^2}{\pi w_i} 
\sum\limits_{n=0} ^{+\infty}
\int \limits_{0}^{+\infty}dk \, I'_{\l}(kr') \frac{A_{n}(k)}{1+\delta_{n,0}} 
\,, &\mbox{if $r'<r_i$}\,, \\
- \frac{q^2 \, L_{\om_e}}{4\pi \lbr r_2(r') \rbr^2}  +
\frac{4q^2}{\pi w_e}
 \sum\limits_{n=0}^{+\infty}
 \int \limits_{0}^{+\infty}
 dk \, K'_{\n}(k r_2 (r')) \frac{B_{n}(k)}{1+\delta_{n,0}} 
\,,  &\mbox{if $r'>r_i$}\, .
\ea \right.
\ee
This is the only force acting over the particle in the locally flat Type I spacetime. To see the results the self-force is plotted for different values of the coupling constant in two distinct Type I geometries; one corresponding to a thin-shell of ordinary matter, i.e. $\k<0$, and the other with a shell of exotic matter, $\k>0$. In the first case, presented in Figure \ref{f1}, the background is a Minkowski interior region ($\om_i=1$) and a conical exterior with $\om_e = 1/2$, while the latter, shown in Figure \ref{f2}, is a conical interior with $\om_i = 1/2$ joined to a Minkowski exterior. The force over the particle is plotted as $\frac{\pi \,{r_i}^2}{4 \,q^2} f$ against the dimensionless position $r/r_i$ of the charge. 

In Figure \ref{f1} we show the results in a Type I geometry with an ordinary matter thin-shell and coupling values in the stable domain $\xi > \xi_c^{(n=1)} = -2$.
The self-force vanishes at the central axis of the geometry because the first term in (\ref{f_I}) is null after regularization in a Minkowski interior. When approaching the shell from either sides we observe a divergent force, being attractive or repulsive depending on the value of the coupling, Figure \ref{f1a}. Nevertheless, in the conical exterior region the force is asymptotically attractive to the center for every value of the coupling constant due to the leading term $\sim - L_{\om_e} r^{-2}$ obtained after renormalization, Figure \ref{f1b}.
    
The solid line in Figure \ref{f1} corresponds to minimal coupling; in the interior region with $r<r_i$ the force increases from cero at the center to infinity as $r \to r_i^-$, while in the exterior we see a negative force diverging at the vicinities of the shell and vanishing at the asymptotic conical infinity. This shows that for $\xi=0$ the curvature jump at the ordinary matter shell acts attracting the particle. We observed that negative values of $\xi$ present qualitatively the same results but with an intensified force near the shell. A richer spectrum appears for positive couplings. If $0< \xi \leqslant 1/4$ the force changes sign in either region but becomes attractive to the shell in its vicinities. For $\xi > 1/4$, the force diverges if $r \to r_i$ but with the opposite sign $i.e.$, the particle is repeled in the vicinities of the ordinary shell. Despite this local difference, the leading asymptotic term proportional to $- L_{\om_e=1/2}$ in (\ref{f_I}) produces an attraction sufficiently far from the shell in the deficit angle exterior region.


In Figure \ref{f2} we show the results in a Type I geometry with a shell of exotic matter and couplings in the stable range $\xi < \xi_c^{(n=1)} = 2$.
The self-force diverges at the center with the particle been attracted to the conical singularity at the interior, and approaching the shell from either sides it diverges showing different behaviors depending on the value of the coupling, Figure \ref{f2a}. In an exterior region without angle deficit $L_{\om_e=1} = 0$, and the force becomes repulsive from the center sufficiently far from the shell for any value of $\xi$, Figure \ref{f2b}, differing from the previous case shown in Figure \ref{f1b} of a deficit angle exterior.

The solid plotted line corresponds to minimal coupling; the force is negative in the interior, $i.e.$ attractive to the conical axis and repulsive from the shell, and in the exterior region the force is positive. For $\xi=0$ the exotic shell repels the particle and the force diverges if $r \to r_i$. Negative values of $\xi$ present similar results but with a greater intensity of the force. The richer spectrum appears, again, for $\xi>0$. With couplings in the range $0< \xi \leqslant 1/4$, the force may vanish and change sign in a same region, but sufficiently near to the shell it becomes repulsive. Oppositely, for $\xi > 1/4$ we see that the shell attracts the particle at its vicinities and sufficiently far from it, in the exterior region, a leading asymptotic repulsion from the center is observed due to the influence of the angle deficit of the core geometry.

In either example we observed a smooth change in the self-force plot as long as the coupling takes values in the corresponding stable domain, with the exception of the change at $\xi = 1/4$.
If the coupling si $\xi_c^{(n=1)} = 1/(\om_e - \om_i)$ we obtain a divergent self-force due to the contribution of the unstable $n=1$ mode of the field. As expected, a  divergency is also encountered at $\xi_c^{(n)}=n/(\om_e - \om_i)$ for $n \in \mathbb{N}$ in the respective mode, corresponding to the critical configurations for Type I geometries.
For the spacetime with an ordinary matter thin-shell (with $\om_i=1$, $\om_e=1/2$ and $\k<0$), the critical couplings are $\xi_c^{(n)}=-2n$ and observing plots for couplings in the successive intervals $-2(n+1)< \xi < -2n $ we found that the self-force is qualitatively the same as the one produced with couplings in the interval $-2<\xi< 0$. Analogously, in the spacetime with exotic matter ($\om_i=1/2$, $\om_e=1$ and $\k>0$), $\xi_c^{(n)}=2n $ and varying the coupling in the intervals $2n < \xi < 2(n + 1)$ we observed qualitatively the same self-force found for couplings in the interval $1/4 < \xi < 2$.


\subsection{Self-force in Type II geometries}
\lb{s4.2}

The regularization procedure over the field obtained in Type II wormhole geometries of radial functions (\ref{radial_II}), for the charged particle placed in $\mathcal{M}_+$ where $r_2(r) = r + r_+$ and $r'_2 = r' + r_+$, yields 
\be\lb{reg_wh}
\P^{R}(r,r') = 
 \frac{q\,L_{\om_+}}{4\pi r'_2}  \lbr 2 - \ln{\frac{r_2(r)}{r'_2}} \rbr + 
\frac{4q}{\pi w_+} \sum\limits_{n=0} ^{+\infty} \int \limits_{0}^{+\infty}dk  \,  K_{\n}(k r_2(r)) \frac{W_{n}(k)}{1+\delta_{n,0}} \,.  
\ee
The self-energy is given by $U_{self}=\frac{q}{2} \P^{R}|_{r=r'}$, and the radial scalar self-force $f= q \, \lp \partial_r \P^R \rp |_{r'}$ is
\be \lb{f_wh}
f =
- \frac{q^2}{4\pi} \frac{L_{\om_+}}{  \lp  r' + r_+\rp^2} + \frac{4q^2}{\pi w_+}
\sum\limits_{n=0}^{+\infty} 
\int \limits_{0}^{+\infty} dk \, K'_{\n}(k (r' + r_+)) \frac{W_{n}(k)}{1+\delta_{n,0}} \,.
\ee

\begin{figure} [h!] 
\centering
\subfloat[Couplings $\xi < 1/4$.]
{\label{f3a}\includegraphics[width=7cm]{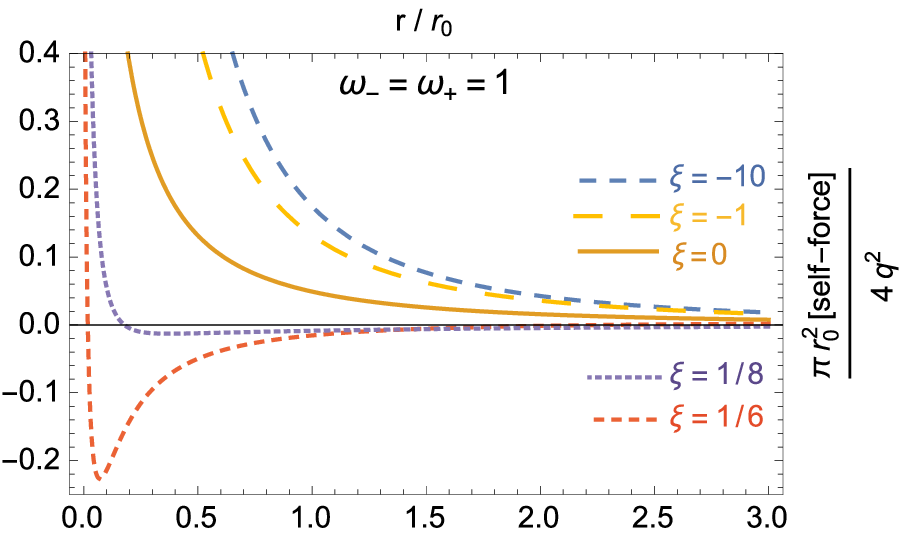}}
\subfloat[Couplings $1/4 <\xi < 1/2 $ and $1/2 < \xi < 1$.]
  {\label{f3b}
    \includegraphics[width=7cm]{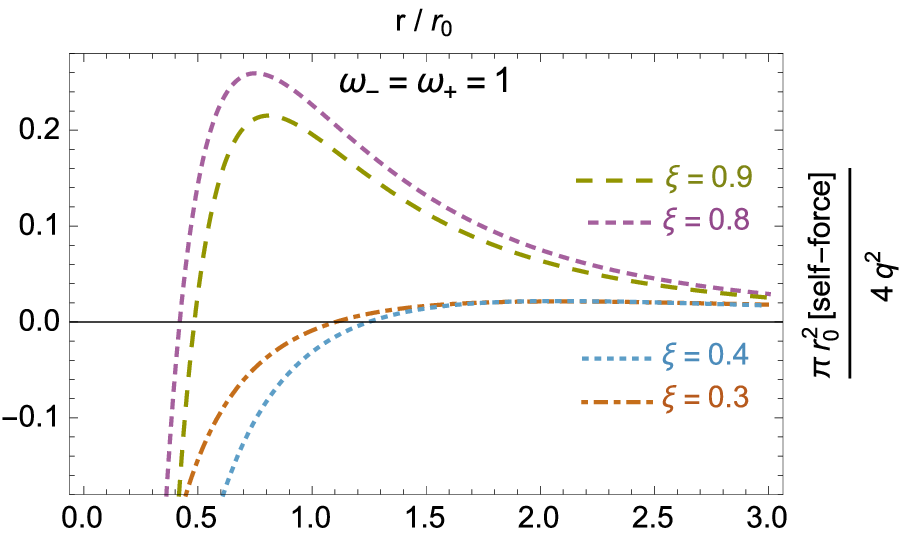}}
    \caption{Dimensionless scalar self-force $\frac{\pi r_0^2}{4q^2} f$ as a function of $r/r_0$ in a Minkowski cylindrical wormhole. The force is asymptotically repulsive.} 
      \label{f3}
    \end{figure}

To analyze the results it is convenient to focus on symmetric wormholes across the infinitely short throat of radius $r_0 = r_+ = r_-$ and, furthermore, we first consider the self-force in a Type II geometry without deficit angle ($\om_+=\om_-=1$). The force over the particle placed in $\mathcal{M}_+$ of a cylindrically symmetric wormhole constructed with two Minkowski regions is plotted in Figure \ref{f3} as $\frac{\pi \,{r_0}^2}{4 \,q^2} f$ against the dimensionless position $r/r_0$ of the charge; couplings $\xi <1/4$ are shown in Figure \ref{f3a}, and couplings $\xi >1/4$ are in Figure \ref{f3b}. In this geometry the first term of the self-force in (\ref{f_wh}) is null, and the common feature is an asymptotically repulsive force from the throat of exotic matter for every value of $\xi$.
The critical configurations for a cylindrically symmetric Minkowski wormhole corresponds to couplings $\xi_c^{(n)}= n/(\om_+ + \om_-) = n/2$, $n \in \mathbb{N}$. An instability is manifested in the $n=1$ mode at the value $\xi_c^{(n=1)} = 1/2$ for which the self-force diverges. While varying the coupling in the safe domain $\xi < 1/2$, the self-force varies smoothly with the exception of the qualitative change produced at $\xi=1/4$. Instabilities are repeated at $\xi_c^{(n)} =  n/2$, and in the successive intervals $n/2 < \xi < (n+1)/2$ we observed qualitatively the same self-force of those cases with couplings $1/4 < \xi < 1/2$; to show this, in Figure \ref{f3b}, we present results for couplings in the interval $1/4 < \xi < 1/2$ and for the next interval, $1/2 < \xi < 1$, as well.

The solid line in Figure \ref{f3a} corresponds to minimal coupling; the force is repulsive from the throat and diverges approaching the shell in the limit $r \to 0$. We see that negative values of $\xi$ produce the same behavior with an increasing repulsion with decreasing value of the coupling. Couplings in the range $0< \xi < 1/4$ show an attractive force in some finite region but produce a repulsion in the vicinities of the infinitely short throat as well. Oppositely, the particle is attracted in the vicinities of the shell  for couplings $\xi > 1/4$, as shown in Figure \ref{f3b},
diverging in the limit $r \to 0$. Despite the different local behavior near the throat for $\xi > 1/4$, we find a leading asymptotic repulsion associated to the positive jump $\k>0$ and the non trivial topology of this spacetime.

\begin{figure} [h!] 
\centering
\subfloat[Couplings $\xi \leqslant 0$.]
{\label{f4a}\includegraphics[width=7.5cm]{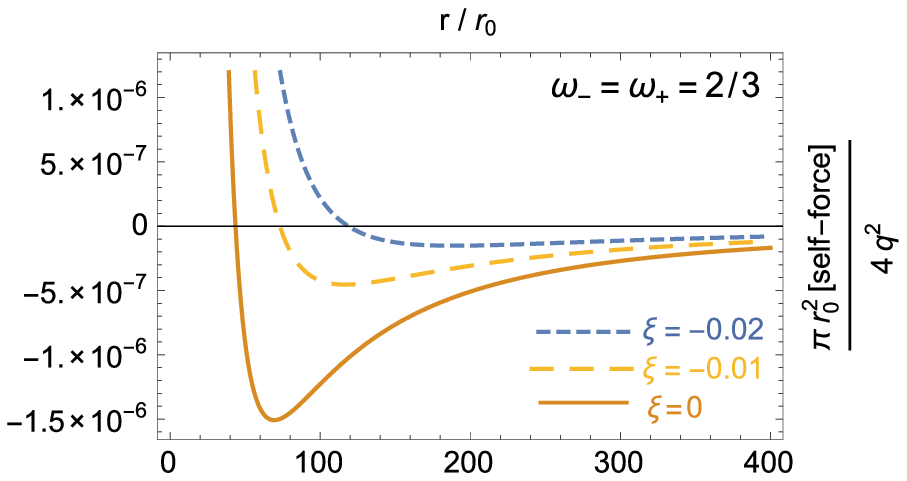}}
\subfloat[Couplings in the range $ 0 < \xi < 1/4$.]
  {\label{f4b}
    \includegraphics[width=6.8cm]{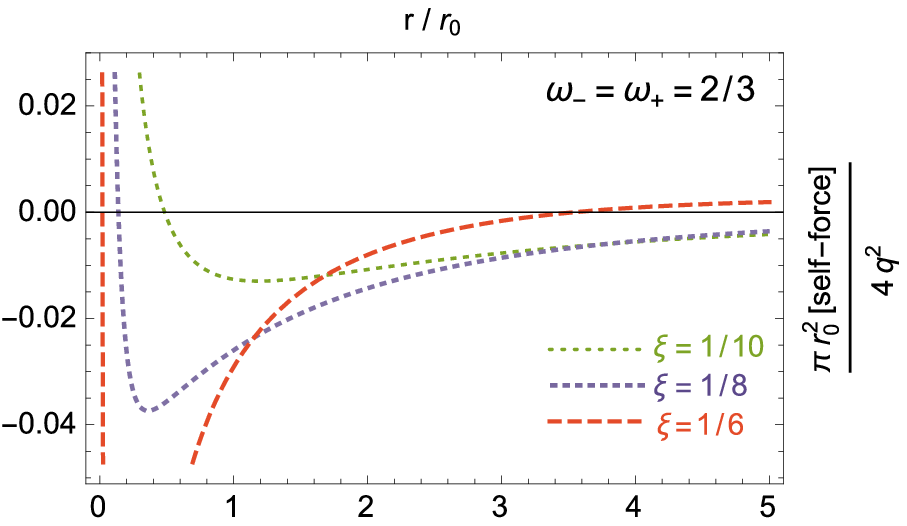}}
    \caption{Dimensionless scalar self-force $\frac{\pi r_0^2}{4q^2} f$ as a function of $r/r_0$ in a conical wormhole. The force is asymptotically attractive.} 
      \label{f4}
    \end{figure}

Finally we present the scalar self-force over a charge placed in a symmetric wormhole constructed with conical regions given by the parameter $\om_+ = \om_- = 2/3$. The force over the particle in $\mathcal{M}_+$ is represented as $\frac{\pi \,{r_0}^2}{4 \,q^2} f$ against the dimensionless position $r/r_0$ of the charge; for couplings $\xi<1/4$ in Figure \ref{f4}, and for couplings $\xi>1/4$ in Figure \ref{f5}. In comparison with the previous Minkowski symmetric wormhole, the main global difference is the asymptotically leading term in (\ref{f_wh}) given by $f \sim - \, L_{2/3}\, r^{-2}$ in the deficit angle regions, which produces an attractive force far from the throat for every value of $\xi$. 
The critical configurations for this conical symmetric wormhole appear for couplings $\xi_c^{(n)}= n/(\om_+ + \om_-) = 3n/4$, $n \in \mathbb{N}$. An instability is manifested in the $n=1$ mode at the value $\xi_c^{(n=1)} = 3/4$ for which the self-force diverges. In the stable domain $\xi < 3/4$, the self-force varies smoothly with the significant qualitative change at $\xi=1/4$, as in previous cases. In Figure \ref{f5} we include cases beyond the stable domain.

The solid line in Figure \ref{f4a} corresponds to minimal coupling; the force is repulsive in the vicinities of the throat, diverges in the limit $r \to 0$ at the thin-shell, and sufficiently far it becomes attractive. Negative values of $\xi$ present the same qualitative behavior with an increasing force with decreasing value of the coupling. Figure \ref{f4b} shows results for couplings in the range $0< \xi < 1/4$; the force is also repulsive near the infinitely thin throat but it may vanish in one or three positions in an intermediate region becoming asymptotically attractive as $r^{-2}$.
In Figure \ref{f5} we show the self-force for couplings $\xi > 1/4$; the force is attractive in the vicinities of the shell for all these cases, we find an intermediate region with a repulsive force and the leading asymptotic attraction associated to the deficit angle term.
We plotted cases with couplings beyond the stable domain; this is $3/4 < \xi < 3/2$, between the critical values for the $n=1$ and $n=2$ modes, which show qualitatively the same self-force of those with couplings in the interval $1/4 < \xi < 3/4$. This last feature is repeated in the successive intervals $3n/4 < \xi < 3(n+1)/4$.
\begin{figure}  [H]
 \centering
    \includegraphics[width=0.75\textwidth]{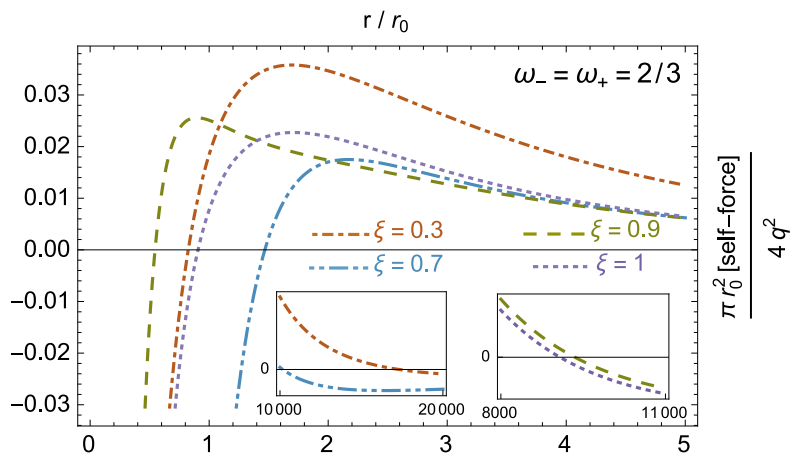}
   \caption{Dimensionless scalar self-force $\frac{\pi r_0^2}{4q^2} f$ as a function of $r/r_0$ in a conical wormhole with couplings in the range $ 1/4 < \xi < 3/4 = \xi_c^{(n=1)}$, and $ 3/4 < \xi < 3/2 = \xi_c^{(n=2)}$. The force is asymptotically attractive.}
      \lb{f5}
\end{figure}



\subsection{The $\xi=1/4$ coupling from the radial equation}
\lb{s4.3}

The analysis of the scalar self-force over a charged particle with the field arbitrarily coupled to the Ricci curvature scalar, in the cylindrically symmetric spacetimes with matter content concentrated over a thin-shell, pointed out the specific value $\xi = 1/4$ as the coupling for which the force changes sign at the vicinities of the shell. The presence of the thin-shell produces a deformation of the field lines and the curvature coupling determines some details of this distortion: in the previously studied examples we saw that the effect over the particle near the shell changes from attraction to repulsion at $\xi=1/4$. 
To understand this result analytically we can study the radial equation (\ref{radial_eq}) by the substitution $\chi_n(k,r) = \psi(r)/\sqrt{\r(r)}$ for each $n$ and $k$, to take it to the form
\be \lb{quantum_eq}  
\lbr \frac{d^2}{d r^2} - V(r,\xi) \rbr \psi(r) = - \,\frac{\delta(r - r')}{\sqrt{\r(r)}} \,.
\ee
The influence of the coupling can be interpreted from the homogeneous version of (\ref{quantum_eq}); a one dimensional radial wave equation for $\psi$ propagating in the potential
\be \lb{potential}
V(r,\xi)=
\frac{4\, n^2 - \r'^2(r)}{4 \r^2(r)} + k^2 + \frac{1}{2} \frac{\r''(r)}{\r(r)} \lp 1 - 4\xi \rp \,,
\ee
where we have used that the Ricci scalar is $R(r)= -2 \, \r''(r)/ \r(r) = - 2 \k  \,\d{(r-r_{s})}$, to put it in terms of the profile function.
The last term in (\ref{potential}), which is proportional to a delta function, changes sign at $\xi=1/4$ in the considered geometries.
Note that in a Minkowski ($\om=1$) or conical spacetime where there is not a thin-shell, this potential is simply
\be 
V_{vac}(r)= \frac{4\, n^2 - \om^2}{( 2r \om)^2} + k^2 \,.
\ee
In comparison with what the radial wave equation would be in a vacuum cylindrical spacetime without the shell, the presence of the matter shell introduces a finite change in $V(r,\xi)$ accounted by $\r'(r)$
in the first term of (\ref{potential}), and an infinitely narrow potential barrier or well at $r=r_s$ given by the delta function on $\r''(r)$. Nevertheless, if $\xi = 1/4$ the \textit{point-like} infinite term disappears from the potential; for this specific value the contribution from the curvature coupling at the thin-shell exactly cancels out the effect of the extrinsic curvature discontinuity in the potential for $\psi$. Without track of the delta function, the potential
\be
V(r,\xi=1/4)=
\frac{4\, n^2 - \r'^2(r)}{4 \r^2(r)} + k^2 		
\ee
has, at most, a finite jump given by $\r'(r)$ (as if it were an interface change at the hypersurface $r=r_s$), and no singularity from the extrinsic curvature discontinuity. 
In terms of a boundary contribution for $\psi(r)$ at the shell, the jump of the radial derivative of $\psi$ at $r=r_s$ can be obtained from
\begin{flalign} \lb{jump_1/4}  
0 & = \int\limits_{r_s-\e}^{r_s+\e} dr \lbr \frac{d^2}{d r^2} - V(r,\xi) \rbr \psi(r) =  \frac{d}{d r} \psi(r)\Big|^{r_s +\e}_{r_s - \e}
- \frac{\k}{2} (1-4\xi)  \psi(r_s)\,.& 
\end{flalign}
While for any value of $\xi \neq 1/4$ the non null jump 
is positive or negative,
for $\xi = 1/4$ there is no jump on $\psi'(r)$ across the shell.\footnote{The value $\xi=1/4$ is known as the coupling that eliminates the Robin boundary energy for a scalar field \cite{fulling2}.}

Focusing in Type I thin-shell spacetimes where we used the global radial coordinate $0<r<+\infty$, the profile function and derivatives are
\be
\r(r) =  \om_i \, r \, \T(r_i - r) + \om_e \lp r - r_i + r_e \rp \T(r-r_i)  \,,
\ee
\be
\r'(r) = \om_i \, \T(r_i - r) + \om_e \, \T(r-r_i)  \,,
\ee
\be
\r''(r) = (\om_e - \om_i) \, \d(r - r_i) = \k \,\r(r_i) \, \d(r - r_i)\,.
\ee
We can see from the potential (\ref{potential}) that the effective infinite term, localized at $r=r_i$, has sign given by the product: $\k(1-4\xi)$. For example, in Type I geometries with ordinary matter thin-shells ($\k < 0$), there is an infinite barrier if $\xi >1/4$ and an infinite well if $\xi<1/4$. In terms of the force produced by the scalar field $\P$ over a charge $q$, in Type I geometries with $\k < 0$; a coupling $\xi > 1/4$ (barrier) repels the particle from the shell in a neighborhood of $r=r_i$, and a coupling $\xi < 1/4$ (well) attracts the particle to the shell. This conclusion is in accordance with the results found in the plots of Figure \ref{f1a} of a shell with ordinary matter. Similarly, the opposite was found in terms of the coupling in the examples plotted on Figure \ref{f2a}; the shell of exotic matter and $\k > 0$ produces an infinite potential with negative sign if $\xi > 1/4$ (well) which attracts the particle, or with positive sign if $\xi<1/4$ (barrier) which repels it, in a neighborhood of $r=r_i$.
In wormhole spacetimes, Type II, the analogous analysis applies; the potential well or barrier with $\xi \neq 1/4$ is manifested in the sign of the self-force at the vicinities of the throat of $\k>0$, producing an attraction ($\xi > 1/4$) or repulsion ($\xi < 1/4$) from the shell.

\section{Summary and conclutions}
\lb{s5}

The arbitrarily coupled massless scalar field produced by a static point charge in cylindrically symmetric backgrounds with a thin-shell of matter was found in spacetimes with one (Type I) or two (Type II) asymptotic regions. The self-force over the charged particle was calculated from the scalar field and studied numerically and analytically in terms of the curvature coupling $\xi$.
The fixed background geometries used are everywhere flat except over the thin-shell where the trace of the extrinsic curvature jump is $\k$ and, for Type I spacetimes only, over the central axes of the geometry in case of conical interiors.
Type I geometries have deficit angle interior and exterior regions, characterized by parameters $\om_i$ and $\om_e$ respectively, with $\k= (\om_e - \om_i)/\r(r_s)$, where $\r(r_s)$ is the metric's profile function at the position of the shell. 
Type II wormhole spacetimes, with exterior regions of parameters $\om_-$ and $\om_+$ respectively, have $\k= (\om_- + \om_+)/\r(r_s)$.
We found the critical values $\xi_c^{(n)} = n/\lp \r(r_s)\,\k \rp$, with $n \in \mathbb{N}$, for which the coupled to curvature scalar field is unstable in the background configuration.
For a Type I geometry with an ordinary matter thin-shell ($\k<0$) the safety domain of the coupling is $\xi > \xi_c^{(n=1)}$, while for exotic matter shells ($\k>0$), either in geometries of Type I or II, the field is stable if the coupling takes values in the range $\xi < \xi_c^{(n=1)}$.
These results add to those in \cite{bezerra,taylor1,taylor2} (mentioned in the introduction) relative to the stable domain of solutions for a scalar field coupled to spherically symmetric wormhole backgrounds but, in our case, studying spacetimes with cylindrical symmetry and in trivial topologies as well.


The only force over the charged particle in these geometries is the scalar self-force, which we obtained from the regularization of its own scalar field.
The sign of the asymptotic force does not depend on the coupling nor on the topological difference between Type I and Type II, it only depends on whether there is an angle deficit or not on the external region where the charge is placed; conical asymptotics produce a leading attractive force,
while Minkowski regions produce a repulsive asymptotic self-force.
This clarify some aspects of the self-force in terms of global properties of the given background geometry. In thin-shell spherical wormholes the scalar self-force of a massless particle changes sign at the value $\xi=1/8$ (corresponding to the conformal flatness of the 3D section of the constant time manifold, see \cite{bezerra,taylor1,taylor2}), in our cylindrical geometries there is no value of $\xi$ for which this occurs globally and the asymptotic sign of the force only depends on the conicity or not of the external region. 
On the other hand, there is a relevant local influence of the coupling over the self-force at the vicinities of the shell of matter. The specific coupling $\xi=1/4$ was identified as the value for which the scalar force changes sign at a neighborhood of the shell; if $\k(1-4\xi)>0$ the shell acts repulsively as an effective potential barrier, while if $\k(1-4\xi)<0$ it attracts the charge as a potential well.
Finally we note that beyond the stable domain of the curvature coupling, in the intervals $\xi \in (\xi_c^{(n)}; \xi_c^{(n+1)})$ between two successive critical values, the self-force shows qualitatively the same behavior as the one produced with couplings in some interval of the safety range; this refers, precisely, to the interval $\xi_c^{(n=1)} < \xi < 0$ in cases with $\k < 0$, or to the interval $1/4 < \xi < \xi_c^{(n=1)}$ in cases with $\k >0$.



\section*{Acknowledgments}
\lb{s6}
This work was supported by the National Scientific and Technical Research Council of Argentina.


\end{document}